\documentclass[journal]{IEEEtran}
\usepackage{amsmath,amsfonts,algorithm,array}
\usepackage[caption=false,font=normalsize,labelfont=sf,textfont=sf]{subfig}
\usepackage{textcomp,stfloats,url,verbatim,graphicx,cite}
\usepackage{soul}
\usepackage{algpseudocodex,circledsteps}
\graphicspath{{figures/}}

\definecolor{G}{RGB}{0,150,0}
\definecolor{B}{RGB}{0,0,150}
\definecolor{R}{RGB}{150,0,0}

\newcommand{\Upd}[1]{#1}
\begin{document}

\title{Split Learning without Local Weight Sharing \\ To Enhance Client-side Data Privacy}

\author{Ngoc Duy Pham, Tran Khoa Phan, Alsharif Abuadbba, Yansong Gao, Doan Nguyen, Naveen Chilamkurti
\thanks{Pham, Phan, Nguyen, and Chilamkurti are with School of Computing, Engineering, and Mathematical Sciences, La Trobe University, Victoria, Australia. Email: {\{ngocduy.pham, k.phan, o.nguyen, n.chilamkurti\}@latrobe.edu.au}.}
\thanks{Abuadbba and Gao are with CSIRO's Data61 \& Cybersecurity CRC, Aus-tralia. Email: {\{sharif.abuadbba, garrison.gao\}@data61.csiro.au}.}
\thanks{{\it{Corresponding authors: T. K. Phan and N. Chilamkurti}}}}

\maketitle

\begin{abstract}
Split learning (SL) aims to protect user data privacy by distributing deep models between the client-server and keeping private data locally. 
In SL training with multiple clients, the local model weights are shared among the clients for local model update. This paper first reveals data privacy leakage exacerbated from local weight sharing among the clients in SL through model inversion attacks. Then, to reduce the data privacy leakage issue, we propose and analyze privacy-enhanced SL (P-SL) (or SL without local weight sharing). We further propose parallelized P-SL to expedite the training process by duplicating multiple server-side model instances without compromising accuracy. Finally, we explore P-SL with late participating clients and devise a server-side cache-based training method to address the forgetting phenomenon in SL when late clients join. The experiment results demonstrate that P-SL helps reduce up to $50\%$ of client-side data leakage, which essentially achieves a better privacy-accuracy trade-off than the current trend by using differential privacy mechanisms. Moreover, P-SL and its cache-based version achieve comparable accuracy to baseline SL under various data distributions, while incurring lower costs for computation and communication. Additionally, caching-based training in P-SL mitigates the negative effect of forgetting, stabilizes the learning, and enables practical and low-complexity training in a dynamic environment with late-arriving clients.
\end{abstract}


\section{Introduction}
Deep learning (DL), influenced by the rapid growth of data, is becoming increasingly important in our daily lives. However, the privacy of data used in the model needs to be protected as required by various privacy regulations \cite{PrivacySurvey21Liu}. Split learning (SL) \cite{OriginalSL18Gupta,SL4Health18Vepakomma,SL4Health19Poirot,SplitLearning22Vepakomma} is one popular collaborative learning technique that aims to protect user privacy by enabling model training without exposing users' raw private data. In a simple vanilla setting, SL divides a deep model into two parts deployed between a client (data owner) and a server (computing service), where only smashed data (local model part's output after feeding raw data) is exposed for collaborative training with the server part \cite{OriginalSL18Gupta}. Compared to federated learning (FL) \cite{FederatedLearning19Yang}, SL is suitable for DL applications on resource-constrained devices (e.g., IoT, mobile) because the clients only need to run the first few layers of the deep model, while the server handles the rest, which involves the most costly computations. With the growing availability of different sources of data, SL has been extended to process learning on multiple clients \cite{OriginalSL18Gupta,CombineSF20Turina,ParallelSL20Jeon,SplitFed22Thapa}. In \cite{EvaluationDL20Gao,EvaluationDL22Gao}, the authors conduct a comprehensive evaluation of SL across various scenarios, ranging from a low to a high number of clients, balanced to imbalanced and extreme data distributions, etc., to provide a thorough insight into SL.

\begin{figure}[h]
    \centering
    \includegraphics[width=0.3\textwidth]{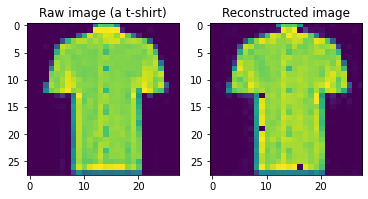}
    \caption{Demonstration of data leakage at the client side of SL: The raw private image (left) is reconstructed (right) by a malicious client through the model inversion attack.}
    \label{fig:leakageSL}
\end{figure}

Regarding SL on multiple data sources, clients typically share their local weights with others to aggregate the learned knowledge from different data sources. This can be done by sequentially passing weights to the next client \cite{OriginalSL18Gupta} or by averaging all local weights at the client side \cite{SplitFed22Thapa}. In these settings, it is assumed that only the server is semi-trustworthy (honest-but-curious \cite{SemiHonest14Paverd}) while all clients trust each other. However, if a client is malicious and colludes with the server, sharing local weights can lead to potential data leakage. Fig. \ref{fig:leakageSL} illustrates an example of data leakage in the original SL \cite{OriginalSL18Gupta} with two clients, $C_1$ and $C_2$. In this scenario, $C_1$ acts as the victim, while $C_2$ is an adversary capable of colluding with the server. $C_2$ employs the model inversion attack \cite{InversionAttack19He} to train a decoder \cite{InversionAttack19He,Autoencoders19Kingma} using $C_1$'s shared local weights. This decoder is then utilized by $C_2$ to reconstruct raw data given $C_1$'s smashed data, which becomes exposed during training or inference as the nature of SL. The smashed data can be acquired by $C_2$ through collusion with the server or eavesdropping on the communication of $C_1$. Furthermore, a decoder trained on $C_2$'s local model could be utilized to attack the subsequent client, which receives $C_2$'s local weights for model updates (e.g., $C_3$ if available). This situation exemplifies the white-box setting of the model inversion attack in \cite{InversionAttack19He}, where the target (local) model is publicly accessible to the nearby/adjacent adversaries due to local weight sharing. Further details about the white-box model inversion attack and its high efficiency can be found in \cite{InversionAttack19He,AttackingProtecting21He}. To address these privacy concerns, we raise the research question (RQ): \textbf{How to develop novel effective SL-based training methods to minimize data leakage in multi-client SL?} In response to this question, we propose a privacy-enhanced SL (P-SL) scheme, which fundamentally obviate sharing local weights among clients during training. The proposed P-SL not only preserves client-server privacy of the original SL but also enhances data privacy at the client side. To the best of our knowledge, this work is the first to identify data leakage in SL exacerbated by the default local weight sharing and investigate P-SL performance under various data distributions.

Furthermore, in SL, apart from the issue of data leakage among clients, ensuring the commitment of all clients to participate in the training process simultaneously poses a significant challenge \cite{EvaluationDL22Gao}. Due to various network, energy, and resource constraints, some devices may not be active throughout the entire training process or may join the training process at a later stage, after collaborative training has already concluded. Handling the training of new clients who join after the initial training, referred to as newcomer training, presents a challenge. This raises another RQ, \textbf{How to ensure stable, low complexity, and high accuracy P-SL in dynamic environments where additional clients join later?} 
As the first training cycle has already been completed, the learning of new clients can deteriorate the knowledge learned by existing clients, leading to the phenomenon of forgetting \cite{Forgetting99French} as recognized in \cite{EvaluationDL22Gao}. To overcome this challenge, we devise a cache-based approach to address the forgetting phenomenon, thus enhancing the learning performance of P-SL. In summary, the contributions of this paper are as follows:

\begin{itemize}
\item \Upd{We consider a stronger adversarial threat model, where all participants (both clients and the server) are assumed to be honest-but-curious}. Based on this threat model, we reveal the issue of client-side data leakage in the original SL and its variants through the lens of model inversion attacks.
\item To address the privacy concerns under the defined threat model, we propose P-SL, which significantly reduces data leakage by up to $50\%$ compared to the original SL. In P-SL, clients no longer share their local weights but can still collaborate with the server to leverage their local knowledge and improve training effectiveness.
\item We conduct a comprehensive empirical evaluation using various datasets with different data distributions to demonstrate that the learning performance of P-SL is comparable to that of the original SL. Additionally, we propose a parallelized version of P-SL that enables simultaneous learning by clients (clients training is performed sequentially in the original SL), reducing the training time without sacrificing accuracy.
\item To tackle the forgetting phenomenon experienced by existing clients when new clients join the training process, we propose a server-side caching approach for P-SL. The approach allows the training of newly arriving clients without the need for retraining existing clients, thereby reducing training overhead. Our experiment results highlight the advantages of caching in SL, particularly in dynamic training environments.
\end{itemize}

The rest of this paper is structured as follows: Section \ref{base} provides background information on SL and its variants, different local data distributions, and the current state of research on privacy preservation in SL. Section \ref{P-SL} presents the identified threat model underlying our proposed P-SL approach. In Section \ref{dynamic}, we explore the parallelization of P-SL and propose the cache-based approach, which handles newly arriving clients to ensure the reliability of P-SL. Section \ref{evaluate} presents the experiment results in terms of accuracy and the privacy preservation capability of the proposal, followed by the conclusion and directions for future research in Section \ref{close}.

\section{Background}\label{base}
This section presents background information on SL with its variants for multiple clients, distribution of user data, and current research on privacy preservation for SL.

\subsection{Vanilla split learning}
A deep model, denoted as $h_\theta:\mathcal{X}\mapsto\mathcal{Y}$, is a hypothesis that maps an input $x\in\mathcal{X}$ to an output $y\in\mathcal{Y}$. Model training involves finding the parameters (weights) $\theta$ that accurately capture the relationship between $\mathcal{X}$ and $\mathcal{Y}$. In order to ensure user data privacy during model training, SL \cite{OriginalSL18Gupta} divides the layers of the deep model into multiple parts. In a simple vanilla setting, the model is split into two components: $h_\theta=f_u\cdot g_w$. The localized part $f_u$ contains the initial layers, while the remaining part $g_w$ is deployed on the server, which is typically the most computationally intensive component. During training, the client performs forward propagation on its local data batch and sends the resulting output (referred to as intermediate data or smashed data) along with the corresponding ground-truth labels $\left(f_u(x^{batch}),y^{batch}\right)$ to the server. The server continues forward propagation on the received smashed data to compute the loss between $y^{batch}$ and $g_w(f_u(x^{batch}))$. The gradients of the loss function are then backpropagated at the server until the split layer, at which point the deep model is cut/split. The split layer's gradients are sent back to the client, where the remaining backpropagation is performed locally, all the way to the first layer. Based on the computed gradients, both the client and the server update their respective weights, $u$ and $w$. This process, known as simple vanilla SL, serves as the core mechanism for many other variants, including SL with multiple clients and our proposed approach.

\subsection{SL with multiple clients}
SL can be extended to train a deep model on $N \ge 2$ clients. The deep model is also split into two parts, $f_\theta=f_u\cdot g_w$, where $f_u$ is distributed to all clients ($f_{u_i}$ to client $C_i$), and $g_w$ is deployed on the central server. The training procedure involves utilizing data from multiple clients in a round-robin fashion. In this setting, when the training process of client $C_{i-1}$ is completed, client $C_i$ receives the weights $u_{i-1}$ of $C_{i-1}$ to initialize its own weights $u_i$. Then, client $C_i$ continues training on its own data, collaborating with the server following the vanilla setting. Once the training is finished, client $C_i$ shares its trained weights, $u_i$, with the next client $C_{i+1}$ \cite{OriginalSL18Gupta}. This process continues until the last client $C_N$ completes training, and the weights $u_N$ trained by client $C_N$ are the model weights that are passed back to all clients for inference.

The model training process in SL is typically performed sequentially among the clients, which can introduce significant latency. To address this issue, several studies have focused on improving the training speed. In \cite{ParallelSL20Jeon}, the authors set up the mini-batch of each client proportional to its local data size, allowing for parallel processing of the training model. All clients are initialized with the exact weights, and after each iteration, the gradients are averaged before being updated on the clients. This synchronization strategy ensures that all clients have the same model weights during training.

SplitFed learning (SFL) \cite{SplitFed22Thapa} is an innovative approach that combines the strengths of FL and SL. In SFL, clients perform forward propagation in parallel on their respective data and send the smashed data to a central server. Upon receiving the gradients from the server, the clients execute the backpropagation step and then send the updated weights to a Fed server. The Fed server aggregates the weight updates using an averaging function ($Avg(\cdot)$) and disseminates a single update to all clients. Similar to \cite{ParallelSL20Jeon}, in SFL, after each global epoch, clients synchronize their models with identical weights, which also renders them susceptible to model inversion attacks in a white-box setting such as an adversarial client which possesses the same model weights as the victims.

\subsection{Privacy-enhancing SL approaches}
Critical privacy vulnerabilities of SL are based on the fact that a neural network is naturally predisposed to be functionally inverted \cite{CombinedFedSplit22Duan}. That is, the smashed data exposed by clients may be exploited to recover the raw input data. Therefore, privacy protection techniques in SL typically aim to minimize data leakage from the smashed data. For example, the noise defense approach \cite{1DCNNSL20Sharif,PracticalDefencesMI21Titcombe} applies additive Laplacian noise to the smashed data before sending it to the server. By introducing noise, the target model is no longer a one-to-one function, making it harder for an attacker to learn the mapping from smashed to raw data. Another method involves adding latent noise through binarization \cite{BinarizingSL22Pham} to reduce the correlation between smashed and input data. However, these mechanisms require efforts to mitigate the impact of noise perturbation on model accuracy \cite{NotJustPrivacy18Ji}.

The work \cite{NoPeek20Vepakomma} aims to reduce raw data leakage by adding an additional distance correlation-based loss term to the loss function. This distance correlation loss minimizes the correlation between the raw and smashed data, ensuring that the smashed data contains minimal information for reconstructing the raw data while still being valuable for achieving model utility. In \cite{BinarizingSL22Pham}, the authors extend this idea by suggesting that the additional loss term can be any leakage metric, not limited to distance correlation. However, the application of an extra loss term may still result in privacy leakage because the smashed data exposes too much information to be adequately protected by a single leakage metric in the loss function \cite{CombinedFedSplit22Duan}. To overcome this limitation, the authors in \cite{FedOrSplit21Turina} propose client-based privacy protection, which employs two different loss functions computed on the client and server sides. In line with this approach, \cite{ResSFL22Li} designs a framework consisting of two steps: a pre-training step that establishes a feature extractor with strong model-inversion resistance, and a resistance transfer step that initializes the client-side models using the feature extractor. This framework requires sufficient computational resources for pre-training on a source task and may be vulnerable during the early training epochs. \Upd{Similarly, \cite{DISCO21Singh} proposes a learnable pruning filter for selectively removing channels in the latent representation space at the split layer (smashed data) to prevent various state-of-the-art reconstruction attacks. However, the pruned network in \cite{DISCO21Singh} may not be suitable for deployment on low-end clients due to the requirement of a large amount of computing power.} To preserve both data privacy and label information,  \cite{MixingActivations22Xiao} employs sample diversity to mix the smashed data of client-side models and create obfuscated labels before transmitting them from clients to the server. The mixed smashed data maintains a low distance correlation with the raw data, thereby preventing private data from being individually reconstructed. However, it should be noted that this mixing technique does not effectively reduce data leakage as intended when performing inference on a single data sample.

In multi-head SL (MHSL) \cite{SplitfedNoSync21Joshi,MultiHeadSL22Joshi}, the authors explore the feasibility of SFL without requiring client-side synchronization. The objective is to reduce the extra communication and computation overhead at the client side using synchronization. The study further considers the case which the server gains information of raw data through the clients' smashed data, but the evaluation does not determine which approach, MHSL or SFL, results in less information leakage. Moreover, the analysis solely focuses on leakage from visualizing smashed data, which exhibits limited quality as demonstrated in \cite{BinarizingSL22Pham}. \Upd{Federated Split Learning (FSL) in \cite{FedOrSplit21Turina} is another study that considers not sharing clients' neural network weights to speed up the training time. The authors propose setting up the same number of clients and edge servers for parallelizing the training with the aim of reducing learning latency.} In contrast, our approach prioritizes privacy and we design the non-local-weight-sharing mechanism to address the privacy attacks under an extended threat model. Additionally, we evaluate our scheme across multiple scenarios and data distributions.

\subsection{SL under diverse data distributions}
In general, data is often distributed among clients in an imbalanced manner. For example, some sensors may be more active than others, resulting more data from certain sources. Similarly, in domains like healthcare, larger institutions tend to have more patient data available \cite{EvaluationDL22Gao}. In a classification task, under balanced data distribution, each client holds samples from all classes in similar quantities. However, in an imbalanced data distribution, each client still possesses samples from all classes, but the total number of samples for each class is imbalanced. It is important to note that the ratio of samples between classes at each client remains similar to the overall dataset ratio. In the study from \cite{EvaluationDL22Gao}, the authors investigate three different data distributions for user data: balanced, imbalanced, and non-IID (non-independent identically distributed). Their findings reveal that SL performs well (compared to FL) under both balanced and imbalanced data while being highly sensitive to non-IID data. Therefore, this study mainly focuses on investigating and evaluating SL specifically under balanced and imbalanced data settings. 

\section{Privacy-enhanced split learning}\label{P-SL}
We define a threat model as the underlying context for the proposed P-SL and the analysis of data leakage.

\subsection{Threat model}
In traditional SL, only the server is assumed to be honest-but-curious \cite{HonestButCurious14Paverd}, meaning it follows the training procedure but may have a curiosity about the raw data from clients. This threat model is assumed in most of the literature on privacy protection techniques in SL. In our study, we extend this assumption to include honest-but-curious clients as well. To the best of our knowledge, our work is the first to consider privacy attacks from both honest-but-curious clients and server in SL.

\noindent{\bf Model Inversion Attack.} Under this identified threat model, any participant in collaborative learning can utilize model inversion attacks to reveal the private data of a target user (client). The model inversion attack consists of three phases: 1) gathering/generating a training set; 2) training a decoder (inverse network \cite{AttackingProtecting21He}) with the data; and 3) recovering raw data from smashed data using the decoder. The attacker can be any adversarial client or the server, as outlined in the following scenarios, with the assumption of client-server collusion for sharing smashed data or querying the local model. Note that due to local weight sharing, all clients in SL have similar local model.
\begin{itemize}
\item \textit{Attacker as a client:} An adversarial client trains the decoder on data generated using its raw data and its local model. Subsequently, the raw data of victim clients can be reconstructed from the smashed data received from the server.
\item \textit{Attacker as the server:} The curious server generates training data by (black-box) querying the adversarial client with a bag of samples (of the same type as raw data \cite{AttackingProtecting21He}). This data is then used to train the decoder, which can be employed to reconstruct raw data from any client's smashed data.
\end{itemize}

We define the leakage of user data as the disparity between the reconstructed data and the original private raw data of a client. The quality of the reconstruction can be evaluated using various metrics such as mean squared error (MSE), structural similarity index measure (SSIM), peak signal-to-noise ratio (PNSR), Kullback–Leibler divergence, and so on \cite{InversionAttack19He,DroppingActivations18Dong,BinarizingSL22Pham,DP4Inference21Ryu}. In our work, we utilize SSIM as the main metric for measuring data leakage.

\subsection{The proposed algorithm}

\begin{figure}[h]
    \centering
    \includegraphics[width=0.4\textwidth]{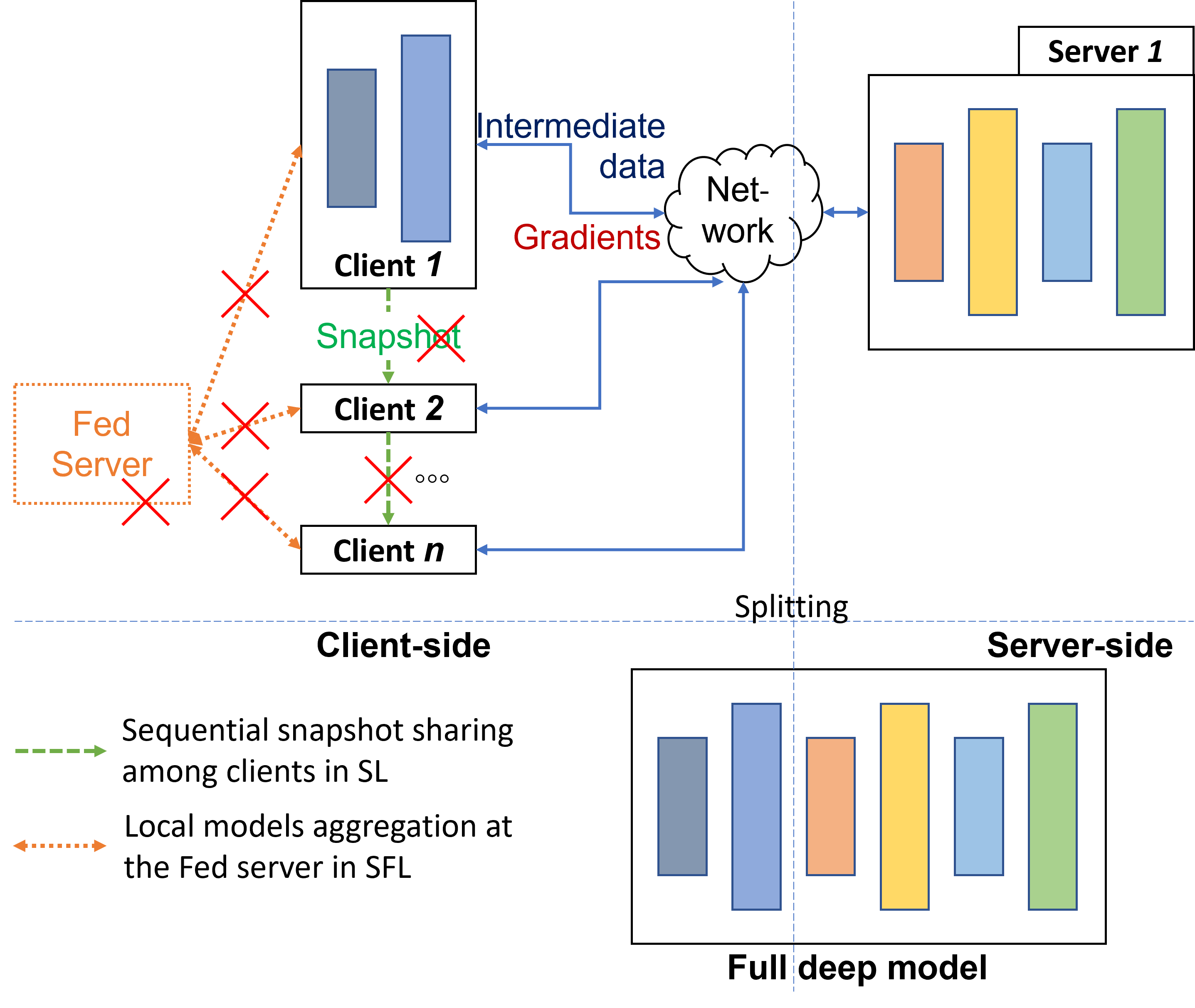}
    \caption{P-SL architecture with differences from original SL \cite{OriginalSL18Gupta} and SFL \cite{SplitFed22Thapa}.
    }
    \label{fig:privacySL}
\end{figure}

To lessen the adverse effects of model inversion attacks, we propose a non-local-weight-sharing method at the client side for SL. The proposed P-SL algorithm is presented in Alg. \ref{alg:privacySL}, followed by computation analysis.

\begin{algorithm}[b]
\caption{Procedure for one global epoch of P-SL.}
\begin{algorithmic}
\State \textbf{Initialize:}
\State $Clients$ and $Server$ receive their model parts
\State $Clients$ and $Server$ initialize their model weights
\end{algorithmic}
\begin{algorithmic}[1]
\For{each $Client_i$ among all the $Clients$}
\While{$Client_i$ has data to train with $Server$}
\State $Client_i$ does forward propagation on its data
\State $Client_i$ sends smashed data and labels to $Server$
\BeginBox
\State $Server$ propagates incoming data on its layers
\State $Server$ computes errors based on the labels
\State $Server$ backpropagates gradients until its first layer
\State $Server$ sends gradients of split layer to $Client_i$
\EndBox
\State $Client_i$ backpropagates the received gradients
\State $Client_i$ and $Server$ update their model weights
\EndWhile
\EndFor
\end{algorithmic}
\label{alg:privacySL}
\end{algorithm}

Fig. \ref{fig:privacySL} illustrates the architecture of P-SL, highlighting its differences from SL and SFL. The proposed P-SL is based on multi-client SL, \textit{where multiple clients connect to a central server, and there is no communication among themselves} (such as sharing snapshots \cite{OriginalSL18Gupta} - local weights) or the use of a Fed server (for local model aggregation \cite{SplitFed22Thapa}). Alg. \ref{alg:privacySL} presents the collaborative training procedure between clients and the server in the proposed P-SL. In the initial phase, clients and the server receive their corresponding parts, $C_i\gets f_{u}$ and $S\gets g_w$, from a split model, $h_{\theta}=f_u\cdot g_w$. Then, they initialize their model weights, $u_i$ and $w$. During a global epoch, following a round-robin manner, each client $C_i$ starts its training with the server, following the simple vanilla SL procedure, which is demonstrated by the inner \textbf{while} loop (lines $2-10$). Note that the box (lines $5-8$) represents the operations executed at the server, and the transmission of data between clients and the server (e.g., smashed data, labels, gradients, etc.) is done via a network connection. Once the training of $N$ clients is completed, \textit{we have $N$ different local models combined with the server model to form $N$ distinct deep models} (i.e. $h_{\theta_i}=f_{u_i}\cdot g_w$ where $1\leq i\leq N$). After training, each client performs inference on its live data using its local private model in combination with the shared server model. In contrast to SL and SFL, P-SL maintains the client-server collaboration but prohibits weight exchanges among clients, thereby reducing local computation, which will be examined in the following.

\subsection{Computation analysis} 
The convergence guarantee of pure SL on homogeneous data is straightforward and can be reduced to the case of SGD \cite{ConvergenceSL23Li}. However, the non-local-weight-sharing of P-SL introduces additional challenges due to the presence of multiple different local models that can be individually combined with the server model, denoted as $h_{\theta_i}=f_{u_i}\cdot g_w$ for training. Let us consider $g_w$ as the primary deep model being trained on the smashed data of all clients. If the training (smashed) data remains stable (with minimal changes), the convergence of P-SL can be simplified to the SGD case. Therefore, the convergence of P-SL relies on the convergence of each client. 
After a few training rounds, when a client's local model has converged, indicating small local weight updates, the corresponding smashed data from that client also stabilizes. 
Stable training data facilitates the convergence of the server-side model's training. Additionally, the learning of the server-side model is influenced by the size of the training data, specifically the size of the smashed data. A larger smashed data size leads to a more extensive training dataset, resulting in a more complex server-side model (e.g., more layers) required to learn and memorize. This insight aligns with the experiment findings in \cite{SplitfedNoSync21Joshi,MultiHeadSL22Joshi}, where the authors evaluate model accuracy with different cutting points (which determine the division of the deep model for deployment on clients and the server, respectively). 
Performance tends to degrade when the client model is thicker (possessing more local layers) because it requires more time to converge, and the corresponding smashed data size also increases. However, for low-end devices, it is preferable to have thin client models, which will facilitate the learning performance of P-SL, as previously discussed.

\begin{table}[b]
\caption{Computation and communication costs at a client of SL, SFL, and P-SL during one global epoch.\label{tab:comp_comm}}
\centering
\begin{tabular}{|l|c|c|}\hline
\textbf{Scheme} &  \textbf{Computation} & \textbf{Communication} \\\hline
SL & $\frac{|\mathcal{X}|}{N}C^P+C^U$ & $2\frac{|\mathcal{X}|}{N}S+2|U|$ \\\hline
SFL & $\frac{|\mathcal{X}|}{N}C^P+C^U$ & $2\frac{|\mathcal{X}|}{N}S+2|U|$ \\\hline
P-SL & $\frac{|\mathcal{X}|}{N}C^P$ \ \ \ \ \ \ \ \ & $2\frac{|\mathcal{X}|}{N}S$ \ \ \ \ \ \ \ \ \ \\\hline
\end{tabular}
\end{table}

To analyze total computation and communication costs of P-SL and provide a comparison to SL and SFL, we assume a balanced data distribution for simplicity. 
Let us consider the following variables: $N$ as the number of clients, $|\mathcal{X}|$ as the total number of dataset items, $S$ as the size of the split layer (the last layer of $f_u$), $C^P$ as the computation cost for processing one forward and backward propagation on $f_u$ with one data item, $C^U$ as the cost for updating a client's local weights from the received weights from the previous client or the Fed server, and $|U|$ as the size of the local model $f_u$. 
Table \ref{tab:comp_comm} demonstrates that the formulated computation and communication costs at the client side in P-SL are lower than SL and SFL, respectively, due to the absence of local weight sharing. The reduction in costs depends on the size of the local model and is independent on data distribution. The factor of $2$ in the communication costs represents the uploading of smashed data and the downloading of corresponding gradients ($2\frac{|\mathcal{X}|}{N}S$), or the uploading and downloading of local models ($2|U|$) at the client side.

\section{P-SL for scalable and dynamic environments}\label{dynamic}
In this section, we further investigate and propose approaches to enhance the performance of P-SL in a dynamic environment where multiple server instances exist or newly participating clients arrive.

\subsection{Parallelizing P-SL with multiple server instances}\label{parallel}
\begin{figure}[t]
    \centering
    \includegraphics[width=0.4\textwidth]{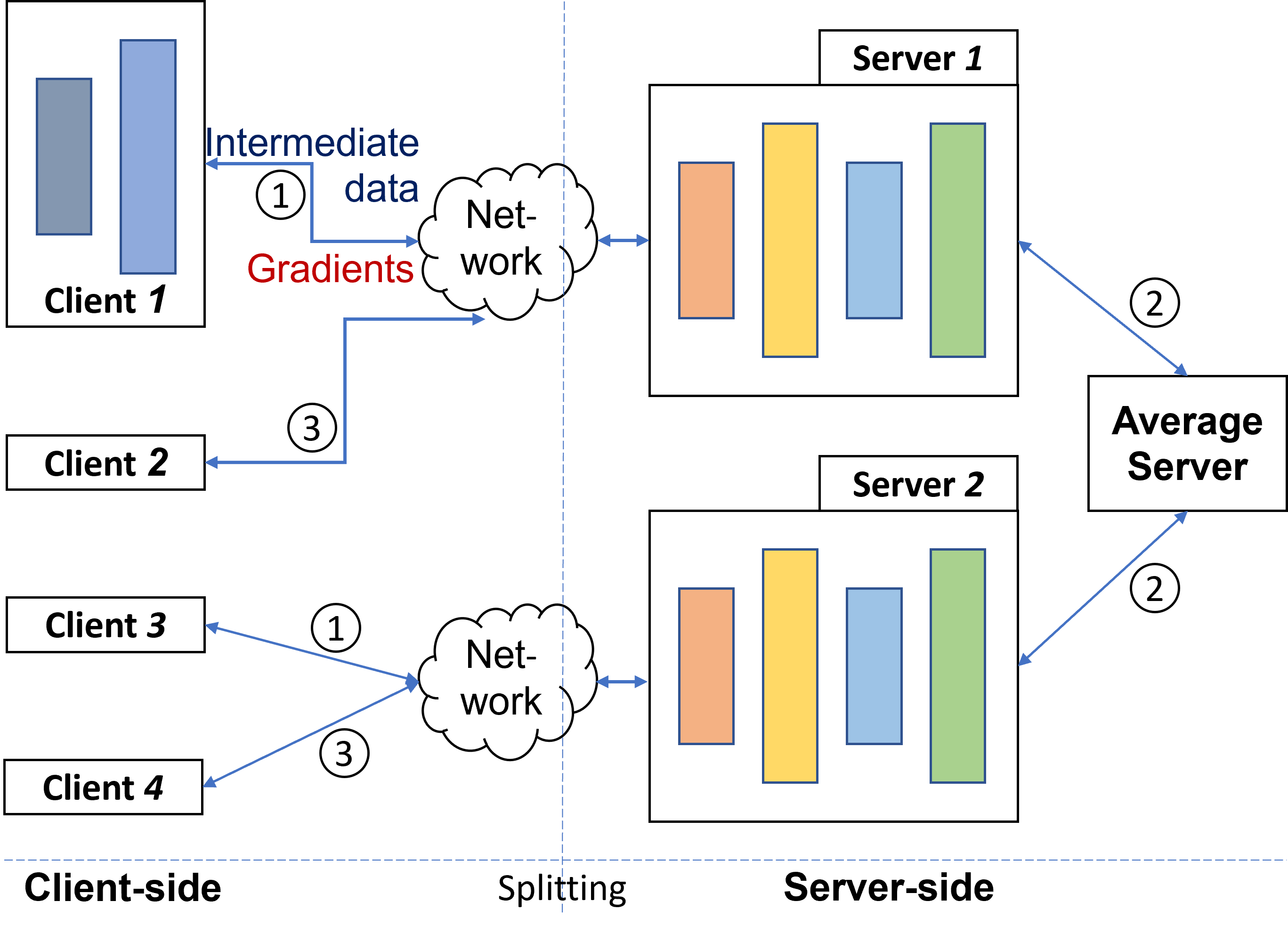}
    \caption{Parallelizing P-SL with two server instances.}
    \label{fig:parallelPSL}
\end{figure}


In the proposed P-SL, each client conducts collaborative training with the server separately. This results in high latency and large idle time at the client side, as only one client is active during training. Therefore, we can process clients' training simultaneously if multiple server instances are available. In Fig. \ref{fig:parallelPSL}, paralleling P-SL follows the steps below in each round:

\begin{enumerate}
\item \textit{Setup phase:} During this phase, all clients receive the same model $f_u$, and the server starts with model $g_w$. The server sets up a pool of $m$ instances (in this example, there are two instances).
\item \textit{Client computation:} Clients connect to the server and are associated with available instances. Then, they perform forward propagation on their local models using their local data in parallel and independently. After this, they send their smashed data to the server.
\item \textit{Server computation:} The corresponding server instances perform forward-backward operations on the received smashed data from the clients and send back the computed gradients.
\item \textit{Client-server collaboration:} The collaborative training between a client and the corresponding server instance is indicated by label \textcircled{1}. Upon completing the training, resulting in a pair client-server model, $f_{u_i}\cdot g_{w_j}$, the server instance becomes available and waits in the pool for the next client to connect (label \textcircled{3}).
\item \textit{Server model aggregation:} When a server instance becomes available after training, a snapshot of the server model weights, $w_j$, is recorded (label \textcircled{2}). After a certain period of time or a predetermined number of snapshots is recorded, the server aggregates (using the $Avg(.)$ function) all snapshots to form a new version of the server model weights, $w^*$. Then all server instances, $g_{w_j}$, update their weights to the new aggregated weights, $w^*$, for the next round of training.
\end{enumerate}

It should be noted that the aggregation of the server models, $g_{w_j}$, is performed asynchronously, and the degree of parallelization depends on the number of server instances. The parallelization of P-SL differs from the client-side parallelization in SFL, as it does not require the Fed server for local model aggregation.

\subsection{P-SL with newly participating clients}\label{cache}
\subsubsection{A case study}
In practice, setting up training when all clients simultaneously participate presents challenges due to the unstable nature of IoT/mobile devices. While previous studies, such as \cite{EvaluationDL22Gao}, examine offline clients' participation during training, there is limited research on the scenario where a new client with its data wants to join the training to benefit from the knowledge acquired by existing clients. To address this real-world situation, we conduct experiments involving $6$ clients, where we initially allow $4$ clients ($C_1,C_3,C_4,C_6$) to collaboratively learn their models using P-SL, referred to as the first training phase. Subsequently, $C_2$ and $C_5$ join the training at a later stage, which we refer to as the second training phase. Both $C_2$ and $C_5$ possess their own data and aim learn their models while leveraging knowledge from the other clients' data. Two possible solutions can be considered for the second training phase: 1. Training all clients, which would impose additional overhead on the existing clients; and 2. Training only the new arriving clients, reducing training complexity. While a hybrid approach, involving training new clients for a few epochs and then training all clients together, is also possible, we focus on the extreme cases (training all or training new) to study the impact of introducing new information to existing knowledge.

\begin{table}[b]
\caption{Accuracy ($\%$) results of $6$ clients, with $2$ joining late\\on the Fashion dataset.\label{tab:new_fashion}}
\centering
\begin{tabular}{|l|c|c|c|c|c|c|}\hline
\multicolumn{1}{|r|}{\textit{Client}} & $\boldsymbol{C_1}$ & $\boldsymbol{C_2}$ & $\boldsymbol{C_3}$ & $\boldsymbol{C_4}$ & $\boldsymbol{C_5}$ & $\boldsymbol{C_6}$ \\\cline{2-7}
\textbf{Training stage} & \multicolumn{6}{c|}{\textit{Balanced data distribution}} \\\hline
\textbf{$1^{st}$} w. 4 clients & $91.4$ & & $91.5$ & $91.6$ & & $91.4$ \\\hline
\textbf{$2^{nd}$} w. ALL clients & $92.6$ & $92.6$ & $92.6$ & $92.4$ & $92.3$ & $92.3$ \\\hline
\textbf{$2^{nd}$} w. NEW clients & $91.0$ & $91.4$ & $90.9$ & $90.6$ & $91.6$ & $91.2$ \\\hline
\textbf{Training stage} & \multicolumn{6}{c|}{\textit{Imbalanced data distribution}} \\\hline
\textbf{$1^{st}$} w. 4 clients & $88.7$ & & $91.0$ & $91.6$ & & $92.1$ \\\hline
\textbf{$2^{nd}$} w. ALL clients & $90.5$ & $90.2$ & $92.3$ & $92.6$ & $92.6$ & $93.2$ \\\hline
\textbf{$2^{nd}$} w. NEW clients & $87.5$ & $89.8$ & $91.0$ & $91.2$ & $92.1$ & $92.1$ \\\hline
\end{tabular}
\end{table}

\begin{figure}[h]
     \centering
     \subfloat[Balanced data]{
        \includegraphics[width=.23\textwidth]{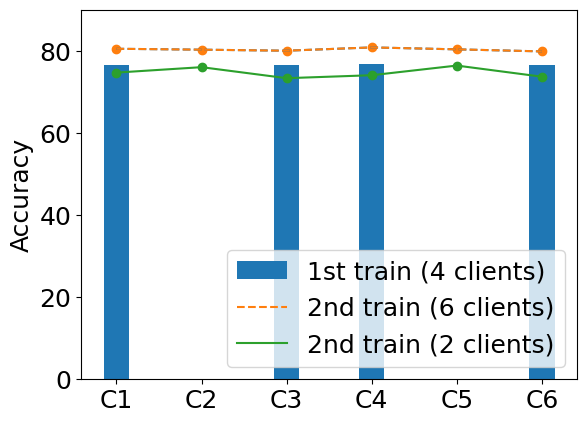}
     }
     \hfill
     \subfloat[Imbalanced data]{
        \includegraphics[width=.23\textwidth]{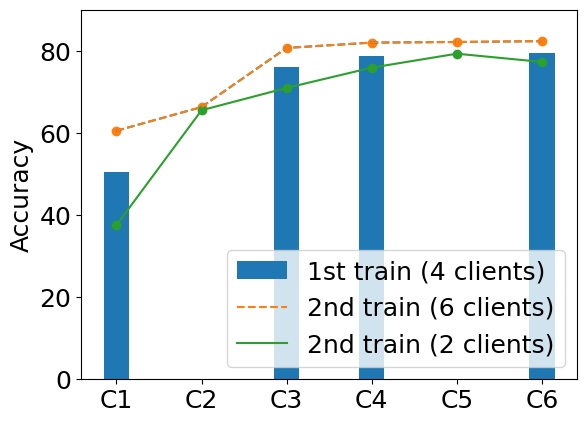}
     }
    \caption{Accuracy ($\%$) results of $6$ clients, with $2$ joinings late on CIFAR10 dataset.}
    \label{fig:new_cifar}
\end{figure}

We conduct experiments on both the Fashion and CIFAR10 datasets to investigate the scenarios involving new clients joining the training process. The detailed settings are deferred to the experiment evaluation section. Table \ref{tab:new_fashion} presents the accuracy results of the first training phase (without $C_2$ and $C_5$), the second training phase with solution one (training all clients), and solution two (training new clients only). Note that all training are performed using P-SL. From the obtained results, we observe that training all clients helps new clients learn their deep models while slightly improving the accuracy of existing clients (e.g., $C_1$ and $C_6$) due to the reinforcement learning from the newcomers' data. Conversely, training only the new clients leads to the server forgetting the knowledge learned from the existing clients, thereby reducing the learning performance of the new joining clients ($C_2$ and $C_5$). Additionally, the accuracy of the existing clients also decreases due to the updating of the server model during training with the new clients. Similar effect can be observed in Fig. \ref{fig:new_cifar}, which visualizes the results on CIFAR10. After the second training phase with the new clients only, the accuracy of the existing clients dropped by $10\%-20\%$ with both balanced and imbalanced data, indicating the phenomenon of forgetting in deep learning. Therefore, training all clients when newcomers join is a suitable approach to maintain the benefits of collaborative learning. However, retraining the existing clients incurs network and computation overhead, which is a limitation for low-end devices.

In summary, our experiments on P-SL involving clients joining after the initial learning phase demonstrate that retraining the entire network is beneficial for newcomers and enhances the performance of existing clients. However, this approach also incurs additional costs for the existing clients, which can be a disadvantage, particularly for low-end devices. In this context, it is preferable that existing clients do not need to retrain, which in turn reduces accuracy due to the forgetting phenomenon.

\subsubsection{Cached-based P-SL algorithm}

\begin{algorithm}[t]
\caption{Server executes in cache-based P-SL.}
\begin{algorithmic}
\State \textbf{Input:} Smashed data and labels from $Client_i$
\State \textbf{Output:} Gradients of the split layer to $Client_i$
\end{algorithmic}
\begin{algorithmic}[1]
\State $Server$ caches $Client_i$'s smashed data and labels
\If{cache pool is not empty}
\State $Server$ randomly selects some smashed data and labels from cache pool
\State $Server$ concatenates the cached smashed data into $Client_i$'s smashed data
\State $Server$ concatenates the cached labels into $Client_i$'s labels
\EndIf
\State $Server$ propagates the concatenated data on its layers
\State $Server$ computes errors based on the concatenated labels
\State $Server$ backpropagates the gradients until its first layer
\State $Server$ slices the gradients based on the split layer's size
\State $Server$ sends the sliced gradients of split layer to $Client_i$
\end{algorithmic}
\label{alg:cachedSL}
\end{algorithm}

To address the issue of forgetting, we propose an enhanced method for training only newcomers to reduce the additional cost of retraining while preserving the knowledge acquired by existing clients. Our approach involves caching the smashed data sent from clients to the server during training, which enhances the learning process of the server model. By caching the data, the server can review knowledge while incorporating new information, mitigating the catastrophic forgetting phenomenon that can occur when a model is serially trained among clients. 
To incorporate caching into the server part of P-SL, we modify the execution as depicted in the box from line $5$ to line $8$ in Alg. \ref{alg:privacySL}. This modification enables the server to cache smashed data from all clients. Subsequently, this cached data can be combined with incoming smashed data during the training of the next client, allowing the server to `review' previous knowledge. The specific details of this modification are presented in Alg. \ref{alg:cachedSL}.

For each iteration of client $C_i$'s training, upon receiving the smashed data, $z_i=f_{u_i}(x_i^{train})$, and the corresponding labels, $y_i^{train}$, the server stores them in a cache pool (line $1$). Before performing forward propagation, the server randomly selects cached data $(z^{cache},y^{cache})$ from the cache pool and concatenates it with the incoming smashed data and labels from client $C_i$ to form $([z_i,z^{cache}], [y_i^{train},y^{cache}])$ as shown in lines $3-5$. Subsequently, the server proceeds with the forward and backward passes using the concatenated data as usual (lines $6-9$). Let $\mathcal{L}$ denote the loss function used to measure the discrepancy between the ground-truth labels and the model's predicted outputs. The gradients at the server's last layer are computed as follows:

\begin{align*}
\nabla\mathcal{L}(\mathrm{outputs},\mathrm{labels})=\\ \underset{u_i,w}{\nabla}&\left[\mathcal{L}\left(g_w([z_i,z^{cache}]),[y_i^{train},y^{cache}]\right)\right]
\end{align*}
It is important to note that the computed gradients for the split layer have the size of the concatenated data instead of the size of $z_i$. Therefore, the server needs to slice the gradients to fit the size of $z_i$ before sending them to the client (line $9$). The execution in clients remains the same as in P-SL, as shown in Alg. \ref{alg:privacySL}. From the above equation, the gradients are computed not only based on the errors from training with $C_i$'s data but also from the other clients' data (cached smashed data and labels). Consequently, by updating $g_w$ using these gradients, the server can simultaneously learn new knowledge from $C_i$'s data and review knowledge previously acquired from other clients' data.

\subsubsection{Computation and privacy analysis}
In cache-based P-SL, we have made modifications only to the server's procedure, ensuring that the cost at the client side remains the same as in P-SL. The additional costs incurred at the server, such as storing cached data and processing concatenation, are considered acceptable because the server is assumed to possess sufficient computing resources to serve multiple clients. Moreover, we have the flexibility to control the size of the cached data, allowing us to adjust the server's performance accordingly. As a result, cache-based P-SL does not increase the cost at the client side, thereby preserving the benefits when applied to IoT/mobile environments.

In terms of data privacy, P-SL already safeguards against sharing local weights among clients, thereby reducing the risk of model inversion attacks by a malicious client. 
Additionally, the caching approach in cache-based P-SL does not violate any privacy concerns, as the cached data is public by default in SL, and clients willingly to share it with the server to derive the utility of the learning process. In summary, cache-based P-SL does not increase the cost at the client side or compromise the privacy of local private data. However, there is an additional overhead in terms of computing and storage resources at the server, which is more feasible to handle compared to low-end devices at the client side. To comprehensively evaluate the learning performance of the proposed scheme, we conduct experiments and present the results in the following section.

\section{Experiment evaluation}\label{evaluate}
To conduct experiments to evaluate the performance of the proposed P-SL, we consider classification tasks on small-scale image datasets using deep models based on 2D-CNN. Table \ref{tab:dataset_model} summarizes of the selected datasets along with the corresponding Very Deep Convolutional Networks (VGG) \cite{VGG14Simonyan}-based deep models. The deep models are divided into two parts: the first two convolutions are deployed at the clients, while the rest of the model resides on the server.

\begin{table}[h]
\caption{Datasets and corresponding deep learning models\label{tab:dataset_model}}
\centering
\begin{tabular}{|l|c|c|c|c|}\hline
\textbf{Dataset} &  \textbf{Input size} & \textbf{Samples} & \multicolumn{2}{c|}{Deep model architecture} \\\cline{4-5}
& & & \textbf{Client side} & \textbf{Server side} \\\hline
Fashion & $1\times28\times28$ & $60,000$ & 2conv & 4conv+1dense \\
CIFAR10 & $3\times32\times32$ & $60,000$ & 2conv & 8conv+1dense \\\hline
\end{tabular}
\end{table}

In these evaluations, we select two datasets: Fashion \cite{Fashion17Xiao} and CIFAR10 \cite{CIFAR09Krizhevsky}, both of which consists of 10 classes and have separate training and testing sets. We distribute a total of $60k$ samples from the train set among $N$ clients and evaluate the learning performance of each client using the same testing set. To simulate an imbalanced data distribution, we assign a varying number of samples to each client following a half bell curve of the standard normal distribution. Table \ref{tab:split_ratio} provides the details of the total number of data samples allocated to each client for the case of $N=6$ clients.

\begin{table}[h]
\caption{Imbalanced data distribution for $6$ clients\label{tab:split_ratio}}
\centering
\begin{tabular}{|l|c|c|c|c|c|c|}\hline
Client index & $\boldsymbol{C_1}$ & $\boldsymbol{C_2}$ & $\boldsymbol{C_3}$ & $\boldsymbol{C_4}$ & $\boldsymbol{C_5}$ & $\boldsymbol{C_6}$ \\\hline
Splitting ratio & $1\%$ & $3\%$ & $9\%$ & $19\%$ & $30\%$ & $38\%$ \\
No. of samples & $600$ & $1.8k$ & $5.4k$ & $11.4k$ & $18k$ & $22.8k$ \\\hline
\end{tabular}
\end{table}

\subsection{P-SL training accuracy}

Using the selected datasets and local data distributions described in the previous section, we implement P-SL with $N=6$ clients and a central server. After training, we measure the learning performance of each client when performing inference on a test set collaboratively with the server ($h_{\theta_i}=f_{u_i}\cdot g_w$). We compare the results with multiple SL, referred to as $m$SL, where we set up $N$ different SL processes between $N$ client-server pairs ($h_{\theta_i}=f_{u_i}\cdot g_{w_i}$). 
The key difference between P-SL and $m$SL is the collaboration at the server side. In $m$SL, there are $m$ distinct server instances, each associated with a different client, and they do not collaboratively aggregate server models. Conversely, in P-SL, we utilize only one server instance, allowing the server model to learn from the smashed data of the entire dataset.
We also include the results of SL and SFL for benchmarking reference in Table \ref{tab:upper_bound}.

\begin{table}[h]
\caption{Benchmarking accuracy ($\%$) of SL and SFL\label{tab:upper_bound}}
\centering
\begin{tabular}{|l|c|c|c|c|}\hline
Dataset & \multicolumn{2}{c|}{\textit{Balanced data}} & \multicolumn{2}{c|}{\textit{Imbalanced data}} \\\cline{2-5}
& \textbf{SL} & \textbf{SFL} & \textbf{SL} & \textbf{SFL} \\\hline
Fashion & $93.7$ & $93.1$ & $93.7$ & $93.4$ \\
CIFAR10 & $85.6$ & $84.2$ & $85.4$ & $84.6$ \\\hline
\end{tabular}
\end{table}

\begin{table}[t]
\caption{Accuracy ($\%$) results with Fashion dataset\label{tab:train1_fashion}}
\centering
\begin{tabular}{|l|c|c|c|c|c|c|}\hline
\multicolumn{1}{|r|}{\textit{Client}} & $\boldsymbol{C_1}$ & $\boldsymbol{C_2}$ & $\boldsymbol{C_3}$ & $\boldsymbol{C_4}$ & $\boldsymbol{C_5}$ & $\boldsymbol{C_6}$ \\\cline{2-7}
Scheme & \multicolumn{6}{c|}{\textit{With balanced data}} \\\hline
\textbf{$m$SL} & $89.5$ & $90.1$ & $89.8$ & $90.0$ & $89.9$ & $90.2$ \\\hline
\textbf{P-SL} & $92.5$ & $92.5$ & $92.4$ & $92.6$ & $92.6$ & $92.6$ \\\hline
Scheme & \multicolumn{6}{c|}{\textit{With imbalanced data}} \\\hline
\textbf{$m$SL} & $78.6$ & $84.9$ & $88.2$ & $90.4$ & $91.1$ & $92.1$ \\\hline
\textbf{P-SL} & $88.8$ & $91.1$ & $92.1$ & $92.8$ & $92.9$ & $92.9$ \\\hline
\end{tabular}
\end{table}

\begin{figure}[h]
     \centering
     \subfloat[Balanced data]{
        \includegraphics[width=.23\textwidth]{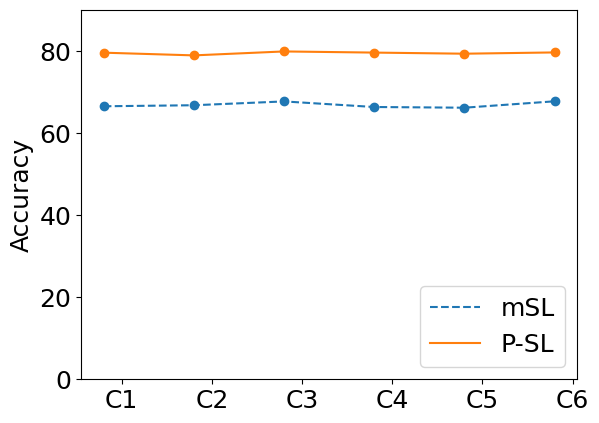}
     }
     \hfill
     \subfloat[Imbalanced data\label{fig:sls_cifar_imb}]{
        \includegraphics[width=.23\textwidth]{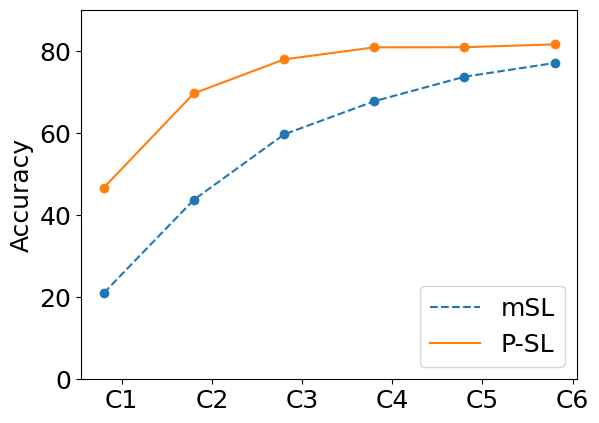}
     }
    \caption{Accuracy ($\%$) results with the CIFAR10 dataset.}
    \label{fig:train1_cifar}
\end{figure}

The training accuracy of each client with the Fashion dataset is presented in Table \ref{tab:train1_fashion}, while the results with the CIFAR10 dataset are visualized in Fig. \ref{fig:train1_cifar}. With $m$SL, the accuracy of each client depends on the number of data samples held by that client, as expected. Therefore, under balanced data, these clients have similar accuracy (around $90\%$ with Fashion), while their accuracy ranges from lower ($C_1$ with $78.6\%$) to higher ($C_{6}$ with $92.1\%$) values with imbalanced data (see Table \ref{tab:train1_fashion}). We can visually observe similar results in Fig. \ref{fig:train1_cifar}, which shows the results using CIFAR10, a more complex and difficult dataset that leads to a higher accuracy difference between clients with fewer and more data samples.

In the proposed P-SL, despite the separate training of local models, the learning performance is better than $m$SL attributing to the shared server model, which aggregates knowledge from all clients. P-SL achieves a $3\%$ higher accuracy with the Fashion dataset and a $12\%$ higher accuracy with the CIFAR10 dataset compared to $m$SL under a balanced data distribution. Under an imbalanced distribution, the results are even more impressive as we observe significant accuracy improvements for clients with fewer data, such as $C_1$, $C_2$, etc. (see Fig. \ref{fig:sls_cifar_imb}). 
This demonstrates the benefit of P-SL in collaborative learning, even without weight sharing among clients. We also compare our results with SL and SFL, which achieve state-of-the-art collaborative learning performance. By sharing local models among clients, knowledge is aggregated at both the client and server sides, resulting in higher accuracy for SL and SFL compared to P-SL, which only aggregates knowledge at the server. In summary, our experiments demonstrate that P-SL, without local weight sharing, still benefits collaborative learning between multiple clients and a central server. Under imbalanced data distribution, clients with less data can learn more by training with clients with more data.

In our experiments, we follow a fixed training order for the clients, starting with $C_1$, followed by $C_2$, and so on up to $C_6$. This fixed training order may have an impact on the accuracy performance since the learning process can be influenced by the presence of more or less data samples during training, especially under an imbalanced data distribution. However, a detailed investigation of the training order and its effects is deferred to the next section for further analysis and discussion.

\subsection{Privacy preservation at client side with P-SL}
We conduct experiments to evaluate the privacy preservation of P-SL. During the training process, we employ model inversion attacks to reconstruct the raw data of all clients using the smashed data that clients send to the server. Specifically, these experiments are conducted with six clients under a balanced data distribution. Whereas all clients are engaged in training, we train a decoder using the local weights and data of one client, which in our case is client $C_1$ (representing a malicious client who is curious about the data of other clients). The trained decoder is used to reconstruct raw data from the smashed data that any client sent to the server. Based on this experiment setup, we evaluate the privacy preservation by measuring the amount of data leakage from clients' raw data. Data leakage is quantified using the SSIM \cite{SSIM04Wang}, which is a perceptual metric that assesses image quality degradation. SSIM provides a measure of similarity between the raw and reconstructed images and has been commonly used in previous works to evaluate data leakage, such as in \cite{InversionAttack19He,DP4Inference21Ryu,SplitUNet22Roth}. Unlike other metrics such as MSE or PSNR, SSIM offers a more intuitive and interpretable metric, where values range between $0$ and $1$. A value of $0$ indicates the least similarity, while a value of $1$ represents the highest similarity, indicating the most leakage.

\begin{figure}[h]
    \centering
    \includegraphics[width=0.49\textwidth]{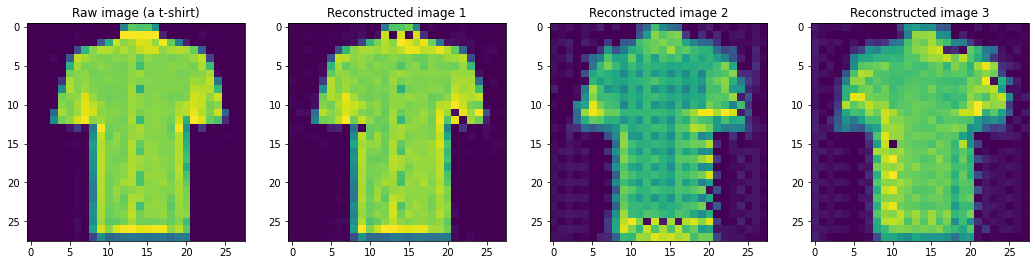}
    \caption{Data leakage at client side in P-SL: raw private image (leftmost) and the reconstructed ones using the smashed data of $C_1,C_2,$ and $C_3$, respectively.}
    \label{fig:leakagePSL}
\end{figure}

Fig. \ref{fig:leakagePSL} demonstrates the reconstructed images from other clients assuming that $C_1$ is malicious. The decoder is trained using $C_1$'s local model and its raw data, resulting in clear reconstructions from $C_1$'s smashed data. However, the quality of reconstruction significantly drops when applying the decoder to the smashed data of other clients. The reconstructed images from $C_2$ and $C_3$ in Fig. \ref{fig:leakagePSL} are vague and contain a high level of noise compared to the raw images. The numerical results presented in Table \ref{tab:leakageIID} reveal that the reconstruction quality (SSIM value) of all clients (except $C_1$) in P-SL is only around $0.5$. On the other hand, with SL and SFL, as partially visualized in Fig. \ref{fig:leakageSL}, $C_1$ can almost entirely reconstruct the raw data of all other clients, with SSIM values exceeding $0.95$. This is similar to $C_1$ self-reconstructing its own data.

\begin{table}[h]
\caption{Data leakage (SSIM) comparison between P-SL, SL,\\and SFL when $C_1$ is the attacker.\label{tab:leakageIID}}
\centering
\begin{tabular}{|l|c|c|c|c|c|c|}\hline
Scheme & $\boldsymbol{C_1}$ & $\boldsymbol{C_2}$ & $\boldsymbol{C_3}$ & $\boldsymbol{C_4}$ & $\boldsymbol{C_5}$ & $\boldsymbol{C_6}$ \\\hline
SL & $0.97$ & $0.96$ & $0.95$ & $0.95$ & $0.95$ & $0.95$ \\
SFL & $0.97$ & $0.97$ & $0.97$ & $0.97$ & $0.97$ & $0.97$ \\
\textbf{P-SL} & $0.97$ & \underline{$0.50$} & \underline{$0.51$} & \underline{$0.52$} & \underline{$0.53$} & \underline{$0.50$} \\\hline
\textbf{P-SL}* & 4$e$-4 & 1$e$-2 & 1$e$-2 & 2$e$-2 & 1$e$-2 & 2$e$-2 \\\hline
\multicolumn{7}{l}{\textit{*: Data leakage is measured using the MSE metric.}}
\end{tabular}
\end{table}

The last row of Table \ref{tab:leakageIID} presents the leakage of P-SL measured by MSE between the raw and reconstructed data. The results show that the errors in reconstructing data for clients $C_2$ to $C_6$ are more than $100\times$ higher than the reconstruction error for $C_1$. While a value of $0$ in the MSE metric indicates full leakage (no reconstruction error), there is no upper bound for non-leak or partial leak scenarios. Additionally, the correlation of leakage among clients does not exhibit a clear and meaningful trend when using MSE, unlike when using SSIM. Therefore, we primarily present the leakage measure using the SSIM metric. In \cite{DP4Inference21Ryu}, the authors made a similar observation and suggested using SSIM to measure leakage instead of MSE and its related PSNR.

\begin{table}[h]
\caption{Data leakage of P-SL under imbalanced data\\with different attackers.\label{tab:leakageImB}}
\centering
\begin{tabular}{|c|c|c|c|c|c|c|}\hline
Attacker & $\boldsymbol{C_1}$ & $\boldsymbol{C_2}$ & $\boldsymbol{C_3}$ & $\boldsymbol{C_4}$ & $\boldsymbol{C_5}$ & $\boldsymbol{C_6}$ \\\hline
$\boldsymbol{C_1}$ & \underline{$0.95$} & $0.32$ & $0.51$ & $0.53$ & $0.48$ & $0.39$ \\
$\boldsymbol{C_6}$ & $0.42$ & $0.45$ & $0.60$ & $0.71$ & $0.73$ & \underline{$0.97$} \\\hline
\end{tabular}
\end{table}

We also conduct experiments with imbalanced data and obtain similar results, as presented in Table \ref{tab:leakageImB}. Based on the experiment results, we can conclude that P-SL outperforms SL and SFL in preserving data privacy at the client side. However, it is important to note that the SSIM values between reconstructed and raw images in P-SL are still significantly different from $0$, indicating the presence of leakage. This leakage can be attributed to the query-free attack described in \cite{InversionAttack19He}, where the attacker does not require knowledge of the target model or the ability to query it. The only assumption for this type of attack is that the attacker and the victim share the same data distribution. In our experiments, the data is uniformly distributed among all clients, regardless of the number of samples they have. Therefore, the local model of the attacker acts as the shadow model for the query-free attack, leading to partial data leakage. Furthermore, an interesting observation from Table \ref{tab:leakageImB} is that attackers with more data (e.g. $C_6$) can reconstruct higher quality images compared to the ones with fewer data (e.g., $C_1$).

\Upd{\noindent{\bf Experiments on 1D time-series data.} Most of the work on SL in the literature typically assumes deep neural network models such as 2D CNN; however, sequential/time-series data are also common in the IoT environment. Therefore, following \cite{1DCNNSL20Sharif}, we conduct our experiments on a 1D time-series dataset to further evaluate the efficiency of our proposed scheme. We set up a two-layer 1D CNN model to classify five heart diseases on the ECG dataset provided by the authors. All other configurations exactly follow \cite{1DCNNSL20Sharif}, except that we split and deploy the first convolutional layer to six clients. After running for $400$ epochs, we successfully reproduce the model training accuracy result, which is $98.4\%$ when using SL. Regarding P-SL, we achieve the comparable model accuracy, which is $97.2\%$. To evaluate the privacy preservation of P-SL, we reconstruct the input data from smashed data and compare the similarity to the raw input using distance correlation (DC) and dynamic time warping (DTW) in alignment with \cite{1DCNNSL20Sharif}.}

\begin{figure}[h]
    \centering
    \includegraphics[width=0.45\textwidth]{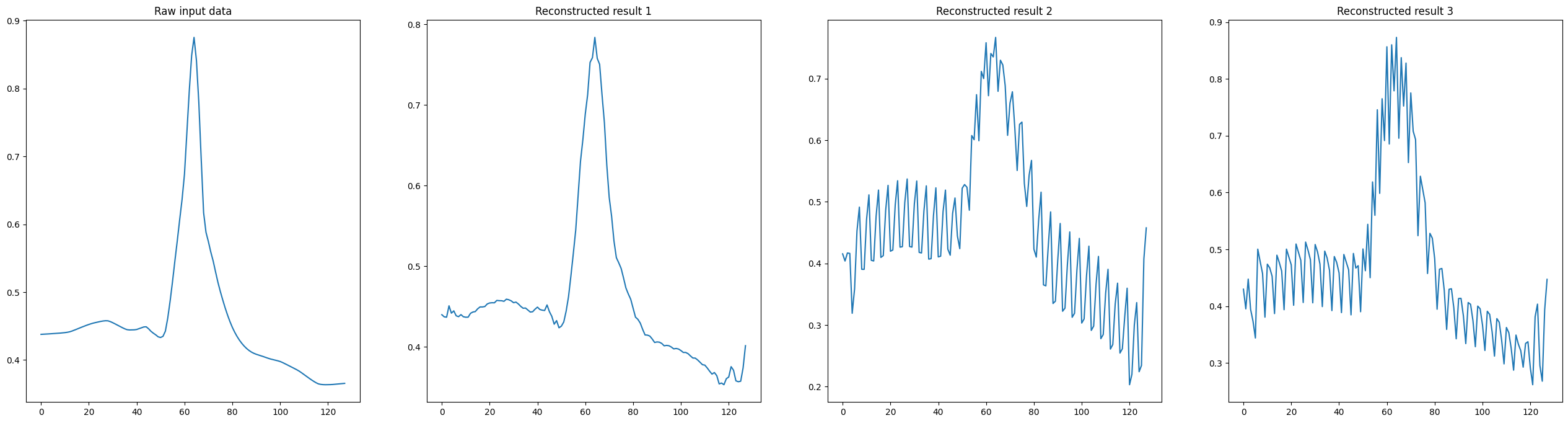}
    \caption{Data leakage at client side in P-SL on ECG dataset: raw input data (leftmost) and the reconstructed results using smashed data of $C_1,C_2,$ and $C_3$, respectively.}
    \label{fig:leakagePSL1D}
\end{figure}

\Upd{Fig. \ref{fig:leakagePSL1D} demonstrates the reconstructed results from other clients assuming that $C_1$ is malicious, regarding experiments on the ECG dataset. Similar to Fig. \ref{fig:leakagePSL}, the decoder trained on $C_1$'s local model and data can effectively reconstruct raw input from $C_1$'s smashed data, with the similarity quantified as $0.15$, $0.99$, and $3e-4$ in terms of DTW, DC, and MSE metrics, respectively. However, the reconstruction quality significantly drops when applied to the smashed data of other clients (i.e. $C_2$ and $C_3$). The mean similarity is $0.69$, $0.86$, and $7e-3$, as measured by DTW, DC, and MSE, respectively. Note that, lower DTW, higher DC, and lower MSE values indicate higher similarity between raw data and the reconstructed results. The experiment results with 1D time-series data further demonstrate the effectiveness of our P-SL in mitigating privacy attacks at the client side of SL.}

\Upd{\noindent{\bf Comparison to related works.} Noise-based protection techniques, such as differential privacy (DP) \cite{DP11Dwork}, are commonly used to ensure guaranteed privacy for user's private data. Recently, various approaches \cite{1DCNNSL20Sharif,LocalDP20Arachchige,DP4Inference21Ryu,BinarizingSL22Pham} have been proposed to apply local DP to protect user data privacy in SL. This makes DP a competitive approach when considering the aforementioned defined threat model. To compare DP to P-SL, we utilize the Laplace mechanism, as described in \cite{DP4Inference21Ryu}. However, it is important to note that there exists a trade-off between accuracy and privacy, as increasing the amount of added noise for higher privacy leads to a reduction in model accuracy. Furthermore, we also experiment with DISCO \cite{DISCO21Singh} and ResSFL \cite{ResSFL22Li}, which are two state-of-the-art approaches aimed at tackling model inversion (MI) attacks by training the client model to be MI-resistant feature extractors. Both of these methods witness accuracy-privacy trade-offs by managing the contribution of resistance factor during training. We quantitatively define privacy as the dissimilarity, represented by $(1-$SSIM$)$, between raw input data and the reconstructed results.}

\begin{figure}[h]
    \centering
    \includegraphics[width=0.25\textwidth]{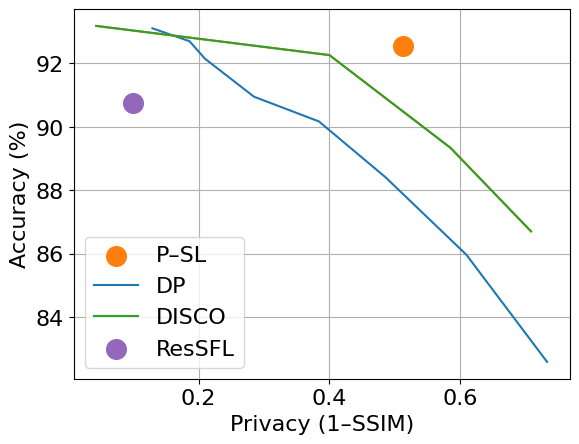}
    \caption{Accuracy - privacy trade-off comparison between P-SL and other related works.}
    \label{fig:tradeoff_DP}
\end{figure}

\Upd{The experiments are conducted on the Fashion dataset with varying noise levels, pruning ratios, and regularization strength of DP, DISCO, and ResSFL, respectively. The results are shown in Fig. \ref{fig:tradeoff_DP}, where we can observe the trade-offs of DP and DISCO (lines) in sacrificing accuracy for improved privacy. On the other hand, our P-SL does not provide a method for controlling the trade-off; however, its result (dot) stays above the lines, indicating that P-SL preserves more privacy (larger dissimilarity in reconstruction) with less accuracy sacrificed. According to the experiments, P-SL achieves privacy similar to applying DP with $\epsilon=1$, while maintaining an accuracy comparable to utilizing DP with $\epsilon=3$. Regarding ResSFL, a simulated attacker is required in training the local model to resist model inversion attacks. However, simulating a weak attacker does not help defend against a strong real attacker, as we can achieve high-quality reconstructed results with a more complex decoder than the one used in training. Furthermore, requiring additional model training (adversarial networks in DISCO and ResSFL) would lead to extra costs for resource-constrained clients in SL. Comparing to other related works, even though lacking flexibility in managing accuracy sacrificed, our P-SL achieves better accuracy-privacy trade-offs with less processing overhead at the client side. Following, we continue to present further evaluation of P-SL such as in dynamic environments.}

\subsection{P-SL in dynamic environments}
\subsubsection{With multiple server instances}
Table \ref{tab:parallel} presents the experiment results of parallelizing P-SL with $6$ clients and $2$ server instances. Each client is randomly associated with a server instance to perform the training. With two servers available, two groups of clients can process training in parallel, theoretically speeding up the training by a factor of two. Based on the reported results, it is evident that parallelized P-SL achieves similar results to sequential P-SL, where we sequentially train each client with a single server. Therefore, parallelized P-SL can be considered `scalable', as it speeds up the training without compromising the model's accuracy.

\begin{table}[h]
\caption{Accuracy ($\%$) results when parallelizing P-SL\\with two server instances.\label{tab:parallel}}
\centering
\begin{tabular}{|l|c|c|c|c|c|c|}\hline
\multicolumn{1}{|r|}{\textit{Client}} & $\boldsymbol{C_1}$ & $\boldsymbol{C_2}$ & $\boldsymbol{C_3}$ & $\boldsymbol{C_4}$ & $\boldsymbol{C_5}$ & $\boldsymbol{C_6}$ \\\cline{2-7}
\textbf{Data dist.} & \multicolumn{6}{c|}{\textit{Fashion dataset}} \\\hline
Balance & $92.1$ & $92.3$ & $92.4$ & $92.2$ & $92.1$ & $92.1$ \\\hline
Imbalance & $90.0$ & $91.0$ & $92.2$ & $92.6$ & $92.8$ & $92.7$ \\\hline
\textbf{Data dist.} & \multicolumn{6}{c|}{\textit{CIFAR10 dataset}} \\\hline
Balance & $81.0$ & $81.6$ & $81.8$ & $81.4$ & $81.4$ & $81.5$ \\\hline
Imbalance & $59.1$ & $74.6$ & $81.1$ & $83.1$ & $84.1$ & $83.8$ \\\hline
\end{tabular}
\end{table}

\Upd{We further evaluate the scalability of P-SL by setting up experiments with $20$ clients connecting to a varying number of server instances. The experiments are conducted on the Fashion dataset, which is distributed evenly among the clients. The results are presented in Table \ref{tab:scalability}, where we report the mean training accuracy of the clients corresponding to the number of server instances set up. Note that the number of server instances corresponds to the speedup in training time. The results exhibit a stable trend in model training accuracy regardless of the number of server instances. Therefore, depending on the resources available at the server side, we can dynamically speed up the training time of P-SL.}

\begin{table}[h]\centering
\caption{Accuracy ($\%$) results of P-SL with $20$ clients and different numbers of server instances.}\label{tab:scalability}
\begin{tabular}{|l|c|c|c|c|c|}\hline
\textbf{Num. of server instances} & 2 & 4 & 5 & 10 & 20 \\\hline
\textbf{Mean accuracy of clients} & $91.1$ & $91.4$ & $91.3$ & $91.1$ & $91.2$ \\\hline
\end{tabular}
\end{table}

\subsubsection{With late joining clients}
\begin{figure*}[t]
\centering
\subfloat[Fashion dataset]{
\begin{tabular}[b]{@{}c@{}}
    \includegraphics[width=.23\textwidth]{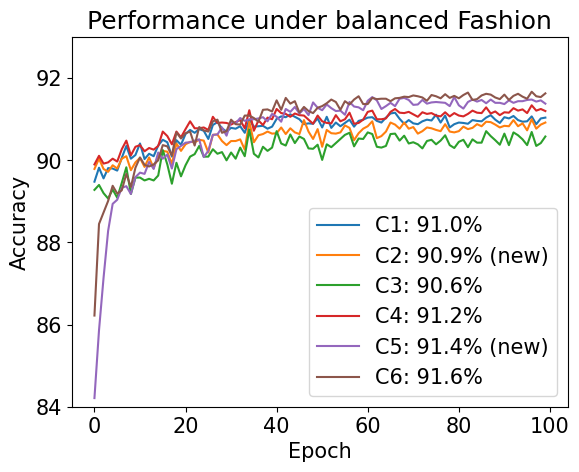}%
    \includegraphics[width=.23\textwidth]{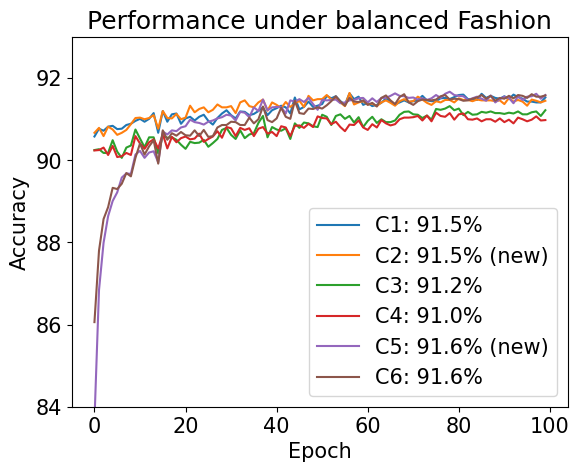}\\
    \includegraphics[width=.23\textwidth]{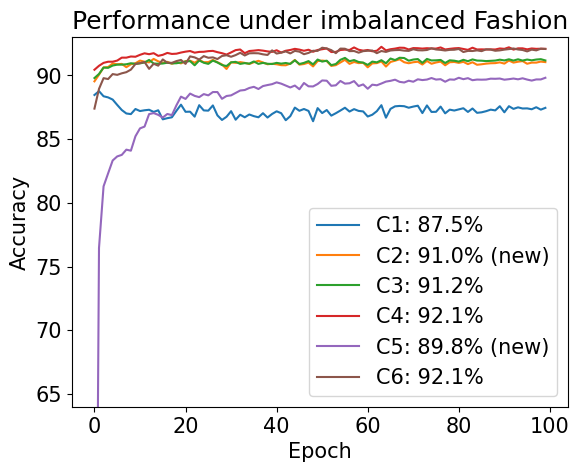}%
    \includegraphics[width=.23\textwidth]{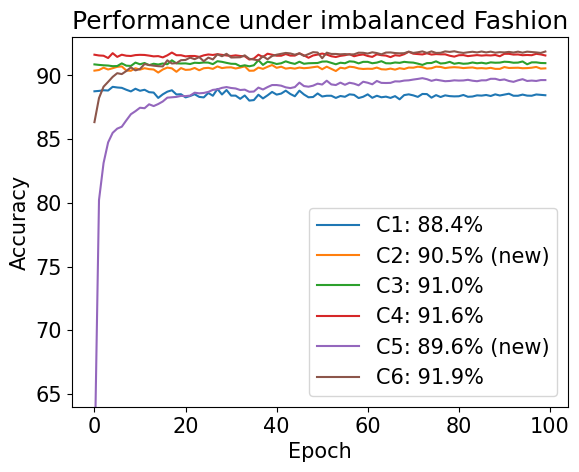}
\end{tabular}}
\hfill
\subfloat[CIFAR10 dataset]{
\begin{tabular}[b]{@{}c@{}}
    \includegraphics[width=.23\textwidth]{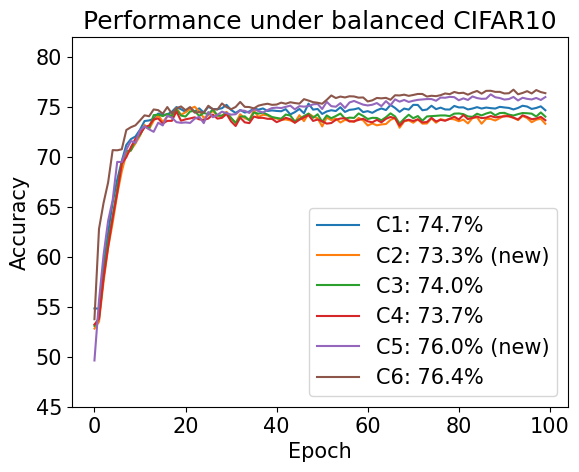}%
    \includegraphics[width=.23\textwidth]{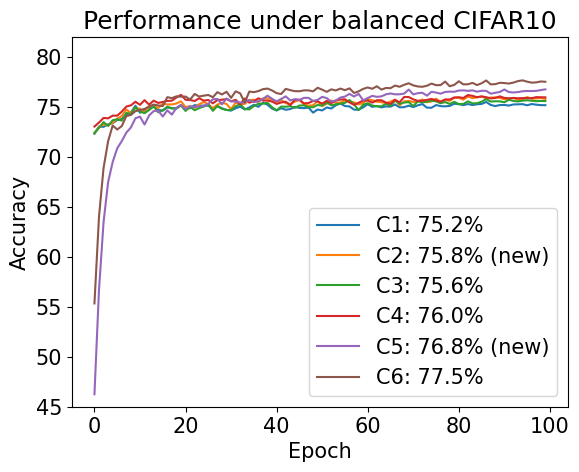}\\
    \includegraphics[width=.23\textwidth]{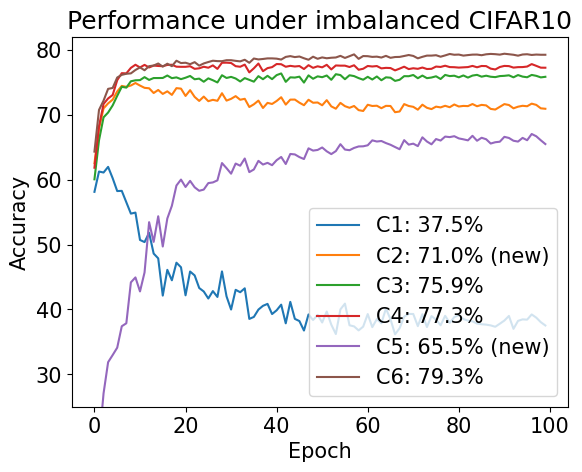}%
    \includegraphics[width=.23\textwidth]{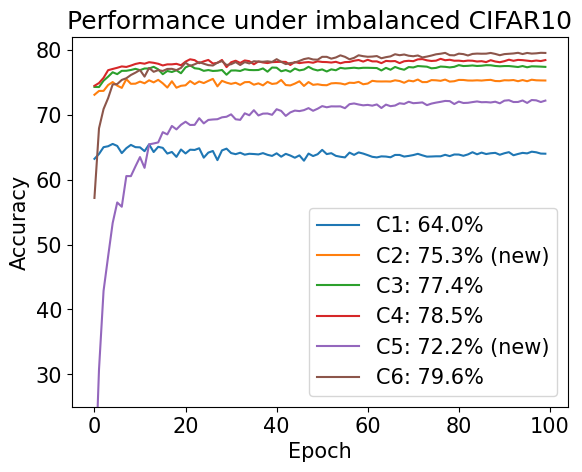}
\end{tabular}}
\caption{Learning performance comparison between non-cached (left column) and cached-based (right column) P-SL with Fashion and CIFAR10 datasets when training newcomers, $C_2$ and $C_5$ (second training).}
\label{fig:new_with_cache}
\end{figure*}

By leveraging cached data during the training of new clients, cache-based P-SL facilitates the review of previous knowledge, resulting in more stable performance and higher accuracy for both new and existing clients. Fig. \ref{fig:new_with_cache} illustrates the learning performance of newcomers ($C_2$ and $C_5$) and existing clients ($C_1$, $C_3$, $C_4$, and $C_6$) using P-SL (left column) and cached-based P-SL (right column) on the Fashion and CIFAR10 datasets. Using caching, learning with only newcomers in P-SL becomes more stable, and the accuracy achieved is comparable to that of retraining the entire network. Therefore, we can train newcomers exclusively using cache-based P-SL, thereby saving the additional costs associated with full retraining while experiencing only a slight reduction in the accuracy of existing clients.

\subsubsection{Order of clients in training}
\begin{figure}[h]
    \centering
    \subfloat[Without caching\label{fig:rand_cifar_imb_noc}]{
        \includegraphics[width=.23\textwidth]{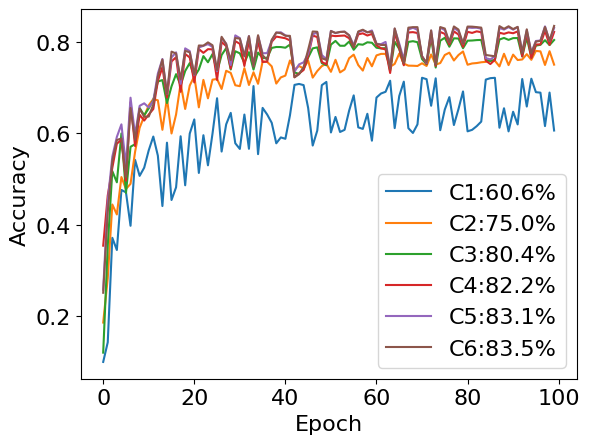}
    }
    \hfill
    \subfloat[With caching\label{fig:rand_cifar_imb_cac}]{
        \includegraphics[width=.23\textwidth]{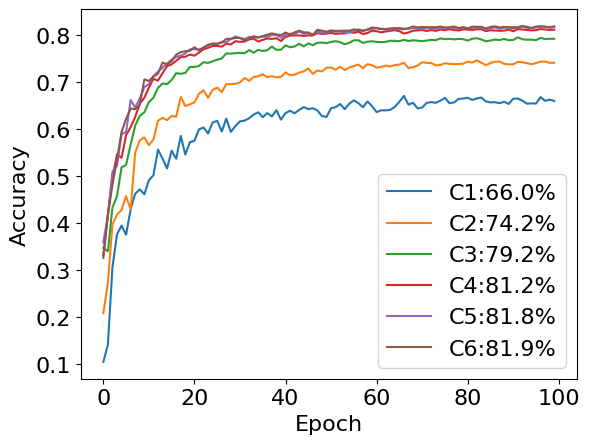}
    }
    \caption{Learning performance of P-SL on imbalanced CIFAR10 where the order of clients participating in the training each epoch is random.}
    \label{fig:random_clients}
\end{figure}

In the previous section, we discussed the impact of the order in which clients participate in the training on the final accuracy, particularly in scenarios with an imbalanced data distribution where some clients have more data than others. We conduct experiments with P-SL, where for each epoch, we randomly select the order of clients to participate in the training with the server. We compare the learning performance to the fixed order, which starts from $C_1$ and ends at $C_6$ each epoch, to assess the effect of client order. The experiment results reveal no significant difference between training with a fixed or random order under a balanced data distribution. Due to the similar quantity and distribution of data, the learning performance of the server with $C_1$ is also similar to that of any other client. However, under an imbalanced data distribution, the learning performance of the server with clients with more data differs from those with fewer data. We plot the learning performance of P-SL in Fig. \ref{fig:rand_cifar_imb_noc}, where it can be observed that the achieved accuracy is not stable. However, when using cache-based P-SL (shown in Fig. \ref{fig:rand_cifar_imb_cac}), the caching approach demonstrates its effectiveness in stabilizing the learning curve.

\subsubsection{Training with Non-IID data}
As the training data collected by the individual clients based on their local environment and usage patterns, it is practical to assume that it is non-IID distributed, e.g., data collected from a single person can only be obtained \cite{FLwithNonIID18Zhao}. We continue to evaluate the learning performance of our proposed method in a non-IID setting, which simulates the worst-case statistical heterogeneity of local data. Following the non-IID setting described in \cite{EvaluationDL22Gao}, we distribute data from only half of the classes ($5$ out of $10$) to each client. The accuracy results obtained after the full training process of SL, SFL, and P-SL with and without caching are reported in Table \ref{tab:caching_nonIID}. The results demonstrate that all the evaluated schemes are sensitive to non-IID data, as the learning objectives of each client diverge when the training data is heterogeneous \cite{EvaluationDL22Gao}. These findings align with those in \cite{EvaluationDL20Gao,EvaluationDL22Gao}, where SL performs worse than SFL in non-IID settings. Our proposed P-SL is particularly sensitive to non-IID data due to the absence of client-side synchronization. Unfortunately, although caching helps stabilize the training loss, it only slightly improves learning accuracy, limiting the applicability of P-SL in non-IID settings. However, caching supports the learning of SL and SFL in non-IID data scenarios, enabling them to achieve comparable results to those obtained in IID data settings (both balanced and imbalanced). This advancement promotes the development of SL in non-IID data environments, an area that has received limited study thus far.

\begin{table}[h]
\caption{Accuracy ($\%$) results of P-SL, SL, and SFL w/w.o. caching under non-IID data distribution.}
\centering
\begin{tabular}{|l|c|c|c|c|c|c|}\hline
& \multicolumn{3}{c|}{\textit{Fashion dataset}} & \multicolumn{3}{c|}{\textit{CIFAR10 dataset}} \\\cline{2-7}
Scheme & \textbf{P-SL} & \textbf{SL} & \textbf{SFL} & \textbf{P-SL} & \textbf{SL} & \textbf{SFL} \\\hline
Without caching & $54.6$ & $54.8$ & $58.2$ & $45.5$ & $52.8$ & $53.8$ \\\hline
With caching & \underline{$56.1$} & \underline{$92.8$} & \underline{$91.2$} & \underline{$47.2$} & \underline{$83.2$} & \underline{$80.0$} \\\hline
\end{tabular}
\label{tab:caching_nonIID}
\end{table}

Based on the above experiments, we can conclude that caching plays a crucial role in stabilizing and maintaining the learning performance of P-SL. As for parallelization to speed up training, this caching approach can be extended to the server with multiple instances that share a cache pool. Furthermore, the strategy for caching, such as determining which and how much data to cache, will be a topic for future research.

\section{Conclusion}\label{close}
This paper addresses the issue of data leakage in traditional SL and its variants that arise from the sharing of local weights during client training. We propose and analyze a variant called SL without local weight sharing, P-SL, to enhance the privacy preservation of user data. The experiment results across various data distributions demonstrate that P-SL enables collaborative learning from distributed clients while reducing data leakage at the client side by half and maintaining comparable accuracy to SL and SFL. Furthermore, P-SL can be parallelized to expedite client training without sacrificing model accuracy. We also investigate P-SL in a dynamic environment where new clients join the training, which can impact the existing clients due to the forgetting phenomenon. To address this, we propose a server-caching mechanism for P-SL, which facilitates the review of learned knowledge during training with newcomers. The experimental results show that cache-based P-SL stabilizes the learning performance and allows for training only the late-arriving clients, reducing client-side overhead while mitigating the server-side forgetting issue. In conclusion, this paper presents P-SL as an effective approach for preserving user data privacy in collaboratively distributed learning, particularly for IoT/mobile devices in real-world dynamic environments. Future research directions include integrating with other privacy-preserving techniques, evaluating on other domains/tasks, and exploring caching strategies to enhance the proposal further.

\bibliographystyle{IEEEtran}
\bibliography{references}

\end{document}